\documentclass[prl,superscriptaddress,showpacs,twocolumn]{revtex4-1}

\usepackage{graphicx}
\usepackage{hyperref}
\usepackage{subfigure}
\usepackage{amsmath}
\usepackage{amssymb}
\usepackage{bbm}

\newcommand{\ket}[1]{| #1 \rangle}

\begin{document}

\title{Implementation of a Toffoli Gate with Superconducting Circuits}
\author{A.~Fedorov}
\author{L.~Steffen}
\author{M.~Baur}
\affiliation{Department of Physics, ETH Zurich, CH-8093 Zurich, Switzerland}
\author{M.P.~da~Silva}
\affiliation{Disruptive Information Processing Technologies Group, Raytheon BBN Technologies, 10 Moulton Street, Cambridge, MA 02138, USA}
\affiliation{D\'epartment de Physique, Universit\'e de Sherbrooke, Sherbrooke, Qu\'ebec, J1K 2R1, Canada}
\author{A.~Wallraff}
\affiliation{Department of Physics, ETH Zurich, CH-8093 Zurich, Switzerland}
%
%

\maketitle

{\bf The Toffoli gate is a three-qubit operation that inverts the state of a target qubit conditioned on the state of two control qubits.  It makes universal reversible classical computation~\cite{Toffoli1980} possible and, together with a Hadamard gate~\cite{Shi2002}, forms a universal set of gates in quantum computation. It is also a key element in quantum error correction schemes~\cite{Cory1998, Knill2001a, Chiaverini2004, Pittman2005, Aoki2009}.  The Toffoli gate has been implemented in nuclear magnetic resonance~\cite{Cory1998}, linear optics~\cite{Lanyon2009}  and ion trap systems~\cite{Monz2009}. Experiments with superconducting qubits  have also shown significant progress recently: two-qubit algorithms~~\cite{DiCarlo2009} and two-qubit process tomography have been implemented~\cite{Yamamoto2010}, three-qubit entangled states have been prepared~\cite{DiCarlo2010, Neeley2010a}, first steps towards quantum teleportation have been taken~\cite{Baur2011} and work on quantum computing architecture has been done~\cite{Mariantoni2011a}. Implementation of the Toffoli gate with only single- and two-qubit gates requires six controlled-NOT gates and ten single-qubit  operations~\cite{Barenco1995}, and has not been realized in any system owing to current limits on coherence. Here we implement a Toffoli gate with three superconducting transmon qubits coupled to a microwave resonator. By exploiting the third energy level of the transmon qubits, we have significantly reduced the number of elementary gates needed for the implementation of the Toffoli gate, relative to that required  in theoretical proposals using only two-level systems. Using full process tomography and Monte Carlo process certification, we completely characterized the Toffoli gate acting on three independent qubits, measuring a fidelity of $68.5\pm0.5$ per cent. A similar approach~\cite{Mariantoni2011a} realizing characteristic features of a Toffoli-class gate has been demonstrated with two qubits and a resonator and achieved a  limited characterization considering only the phase fidelity. Our results reinforce the potential of macroscopic superconducting qubits for the implementation of complex quantum operations with the possibility of quantum error correction schemes~\cite{Reed2011}.}

\begin{figure}[!ht]
  \includegraphics[width= 1.0 \columnwidth]{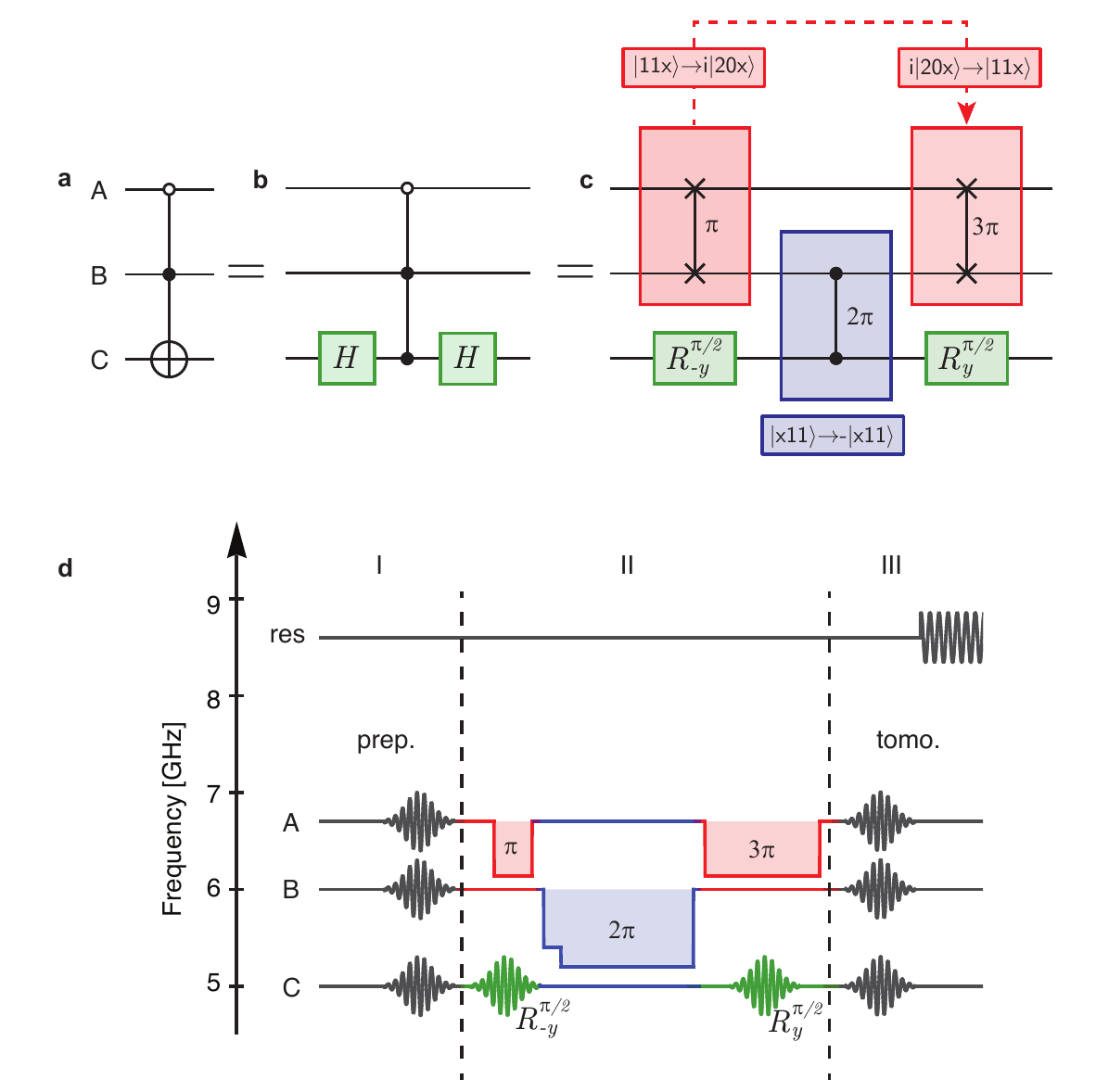}
  \caption{{\bf Circuit diagram of the Toffoli gate.} {\bf a,} A \textsc{not}-operation ($\oplus$) is applied to qubit C if the control qubits (A and B) are in the ground ($\circ$) and excited state ($\bullet$) respectively. {\bf b,} The Toffoli gate can be decomposed into a CCPHASE gate  sandwiched between Hadamard gates (${\bf H}$) applied to qubit C.
  {\bf c,} The CCPHASE gate is implemented as a sequence of a qubit-qutrit gate, a two-qubit gate and a second qubit-qutrit gate. Each of these gates is realized by tuning the $\ket{11}$ state into resonance with $\ket{20}$ for a $\{\pi,2\pi,3\pi\}$ coherent rotation respectively. For the Toffoli gate, the Hadamard gates are replaced with $\pm \pi/2$ rotations about the $y$ axis (represented by $R_{\pm y}^{\pi/2}$).
    {\bf d,} Pulse sequence used for the implementation of the Toffoli gate. During the preparation (I), resonant microwave pulses are applied to the qubits on the corresponding gate lines. The Toffoli gate (II) is implemented with three flux pulses and resonant microwave pulses (colour coded as in {\bf c}). The measurement (III) consists of microwave pulses that turn the qubit states to the desired measurement axis, and a subsequent microwave pulse applied to the resonator is used to perform a joint dispersive read-out.}
\label{fig:circuit}
\end{figure}

We have implemented a Toffoli gate with three transmon qubits (A,B and C) dispersively coupled to a microwave transmission-line resonator, in a sample which is identical to the one used in ref.~\cite{Baur2011}. The resonator is used for joint three-qubit read-out by measuring its transmission~\cite{Filipp2009b}. At the same time, it serves as a coupling bus for the qubits~\cite{Majer2007}. The qubits have a ladder-type energy level structure with sufficient anharmonicity to allow individual microwave addressing of different transitions.  We use the first two energy levels as the computational qubit states, $\ket{0}$ and $\ket{1}$, and use the second excited state, $\ket{2}$, to perform two-qubit and qubit-qutrit operations (a qutrit is a quantum ternary digit). From spectroscopy, we deduce a bare resonator frequency  $\nu_{\mathrm{r}} = 8.625\,\mathrm{GHz}$ with a quality factor of $3300$; maximum qubit transition frequencies $\nu^\mathrm{max}_\mathrm{A} = 6.714\,\mathrm{GHz}$, $\nu^\mathrm{max}_\mathrm{B} = 6.050 \,\mathrm{GHz}$ and $\nu^\mathrm{max}_\mathrm{C} = 4.999\,\mathrm{GHz}$; and respective charging energies $E_c/h = 0.264$, $0.296$ and $0.307\,\mathrm{GHz}$ ($h$, Planck's constant) and qubit-resonator coupling strengths $g/2\pi = 0.36,\,0.30$ and $0.34\,\mathrm{GHz}$ for qubits A, B and C. At the maximum transition frequencies, we find respective qubit energy relaxation times  of $T_1 = 0.55,\,0.70$ and $1.10\,\mathrm{\mu s}$ and phase coherence times of $T_2^* = 0.45,\,0.6$ and $0.65\,\mathrm{\mu s}$ for qubits A, B and C.

In the conventional realization of the Toffoli gate, a \textsc{not} operation is applied to the target qubit (C) if the control qubits (A, B) are in the state $\ket{11}$. In our set-up it is more natural to construct a variation of the Toffoli gate shown in Fig.~1a in which the  state of the target qubit is inverted if the control qubits are in $\ket{01}$. This gate can easily be transformed to the conventional Toffoli gate by a redefinition of the computational basis states of qubit A or by applying two $\pi$-pulses on qubit A.

The Toffoli gate can be constructed from a `controlled-controlled-phase' (CCPHASE) sandwiched  between two Hadamard gates acting on the target qubit as shown in Fig.~1b. A CCPHASE gate leads to a phase shift of $\pi$ for state $\ket{1}$ of the  target qubit if and only if the control qubits are in state $\ket{01}$. In other words, this corresponds to a sign change of only one of the eight computational three-qubit basis states: $\ket{011} \leftrightarrow - \ket{011}$.

\begin{table}[!b]
\caption{List of states after each step of the CCPHASE gate. The state $\ket{011}$ acquires a phase shift of $\pi$ during the CPHASE pulse; the states $\ket{11x}$ are transferred to $i\ket{20x}$, `hiding' them from the CPHASE gate; and the initial states $\ket{x0y}$ and $\ket{010}$ do not change during the sequence.}
\begin{tabular}{|l|l|l|l|}
\hline
Initial state & After $\pi$-SWAP & After CPHASE & After $3\pi$-SWAP\\
\hline
$\ket{011}$&$\ket{011}$&$-\ket{011}$&$-\ket{011}$\\
$\ket{11x}$&$i\ket{20x}$&$i\ket{20x}$&$\ket{11x}$\\
$\ket{x0y}$&$\ket{x0y}$&$\ket{x0y}$&$\ket{x0y}$\\
$\ket{010}$&$\ket{010}$&$\ket{010}$&$\ket{010}$\\
\hline
\end{tabular}
\end{table}

The basic idea of `hiding' states by tranforming them into non-computational states to simplify the implementation of a Toffoli gate was theoretically proposed in refs.~\onlinecite{Ralph2007,Borrelli2011a} and has been experimentally implemented for linear optics and ion trap systems~\cite{Monz2009,Lanyon2009}. The implementation of the scheme of ref.~\cite{Ralph2007} in our set-up would require three controlled-phase (CPHASE) gates, six single-qubit operations and two single-qutrit operations. Instead, we construct the CCPHASE from a single two-qubit  CPHASE gate and two qubit-qutrit gates. The latter gates are called $\pi$-SWAP and 3$\pi$-SWAP, respectively (Fig.~1c, red frames).
The application of a single CPHASE gate to qubits B and C (Fig.~1c, blue frame) inverts the sign of both $\ket{111}$ and $\ket{011}$. To create the CCPHASE operation, the computational basis state $\ket{111}$ is transferred to the non-computational state $i\ket{201}$ by the $\pi$-SWAP gate, effectively hiding it from the CPHASE operation acting on qubits B and C. After the CPHASE operation, $\ket{111}$ is recovered from the non-computational level $i\ket{201}$ by the $3\pi$-SWAP gate. Alternative approaches using optimal control of individual qubits for implementing a Toffoli gate in a single step have been  proposed~\cite{Sporl2007} and recently analyzed in the context of the circuit quantum electrodynamics architecture~\cite{Stojanovic2011}.

All three-qubit basis states show three distinct evolution  paths during our CCPHASE gate (Table 1). Only input state $\ket{011}$ is affected by the CPHASE gate acting on qubits B and C, which transfers $\ket{011}$ to the desired state, $-\ket{011}$. The states $\ket{11x}$ with $x\in \{0,1\}$ are transferred by the $\pi$-CPHASE gate to the states $i\ket{20x}$. The subsequent CPHASE gate therefore has no influence on the state. The last gate ($3\pi$-CPHASE) transfers $i\ket{20x}$ back to $\ket{11x}$.  Together the two SWAP gates realize a rotation by $4\pi$, such that the state $\ket{11x}$ does not acquire any extra phase relative to the other states. The states of the last group ($\ket{010}$ and  $\ket{x0y}$ with $y\in \{0,1\}$) do not change during the CPHASE gate sequence.

The actual experimental implementation of the Toffoli gate consists of a sequence of microwave and flux pulses applied to the qubit local control lines (Fig.~1d). The arbitrary rotations  about the $x$ and $y$ axis~\cite{Gambetta2011a} are realized with resonant microwave pulses applied to the open transmission line at each qubit. We use 8-ns-long, Gaussian-shaped DRAG-pulses~\cite{Motzoi2009, Gambetta2011a} to prevent population of the third level and phase errors during the single-qubit operations. Few-nanosecond-long current pulses passing through the transmission lines next to the superconducting  loops of the respective qubits control the qubit transition frequency realizing $z$-axis rotations.
All two-qubit or qubit-qutrit gates are implemented by tuning a qutrit non-adiabatically to the avoided crossing between the states $\ket{11x}$ and $\ket{20x}$ or, respectively, $\ket{x11}$ and $\ket{x20}$ (refs~\onlinecite{Strauch2003, DiCarlo2010, Haack2010}). During this time, the system oscillates between these pairs of  states with respective frequencies $2 J_{11,20}^{AB}$ and $2 J_{11,20}^{BC}$. With interaction times $\pi/(2 J_{11,20}^{AB}) = 7$~ns, $3\pi/(2 J_{11,20}^{AB}) = 21$~ns and $\pi/(2 J_{11,20}^{BC}) = 23$~ns, we realize a $\pi$-SWAP and a 3$\pi$-SWAP between qubits A and B and a CPHASE gate between qubits B and C, respectively. Our use of qubit-qutrit instead of single-qutrit operations allows for a more efficient construction of the Toffoli gate. Direct realization of the scheme proposed in ref.~\onlinecite{Ralph2007} in our system would require eight additional microwave pulses (used to implement six single-qubit and two single-qutrit gates) with a twofold increase in overall duration of the pulse sequence with respect to our scheme.

\begin{figure}
 \includegraphics[width=1.0 \columnwidth]{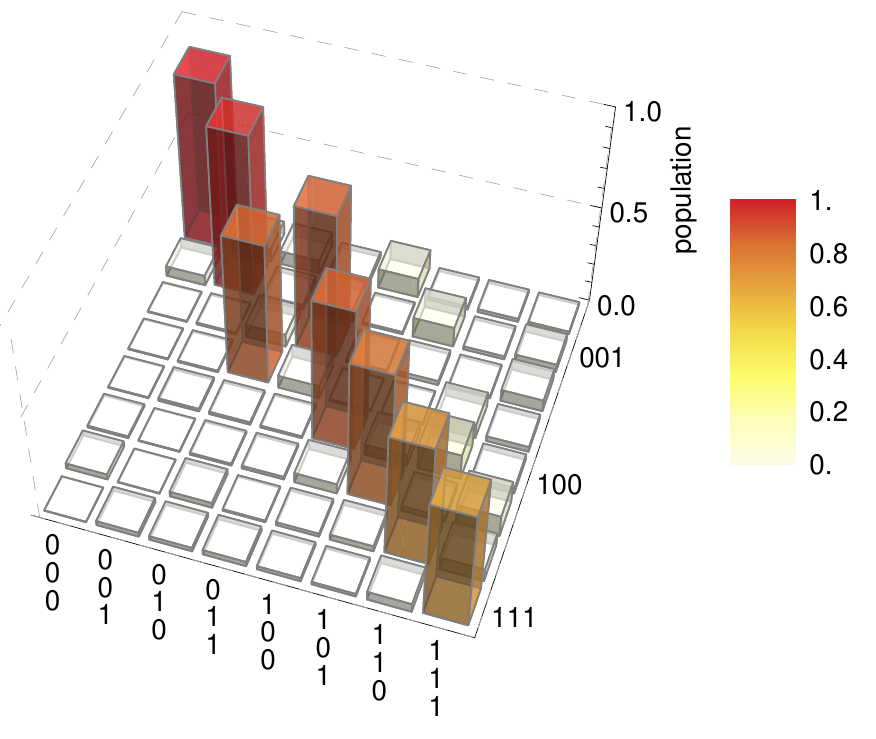}
  \caption{{\bf Truth table of the Toffoli gate.} The state of qubit C is inverted if qubits A and B are in the state $\ket{01}$. The fidelity of the truth table is $F = (1/8)\mathrm{Tr}\left[{U_\mathrm{exp}U_\mathrm{ideal}}\right] = 76.0 \%$}
  \label{fig:truthtable}
\end{figure}

We have characterized the performance of this realization of a Toffoli gate by measuring the truth table, by full process tomography~\cite{Chuang1997} and by Monte Carlo process certification~\cite{Silva2011,Flammia2011}.
The truth table depicted (Fig.~2) shows the population of all computational basis states after applying the Toffoli gate to each of the computational basis states. It reveals the characteristic properties of the Toffoli gate, namely that a NOT operation is applied on the target qubit (C) if the control qubits (A and B) are in the state $\ket{01}$. The fidelities of the output states show a significant dependence on qubit lifetime. In particular, input states with qubit A  (with the shortest lifetime) in the excited state generally have the worst fidelity, indicating that the protocol is mainly limited by the qubit lifetime. The fidelity of the measured truth table, $U_{\rm exp}$, with respect to the ideal one, $U_{\rm ideal}$, namely $F = (1/8)\mathrm{Tr}\left[{U_\mathrm{exp}U_\mathrm{ideal}}\right] = 76.0 \%$, shows the average performance of our gate when acting onto the eight basis states.

\begin{figure*}
  \includegraphics[width=2.1 \columnwidth]{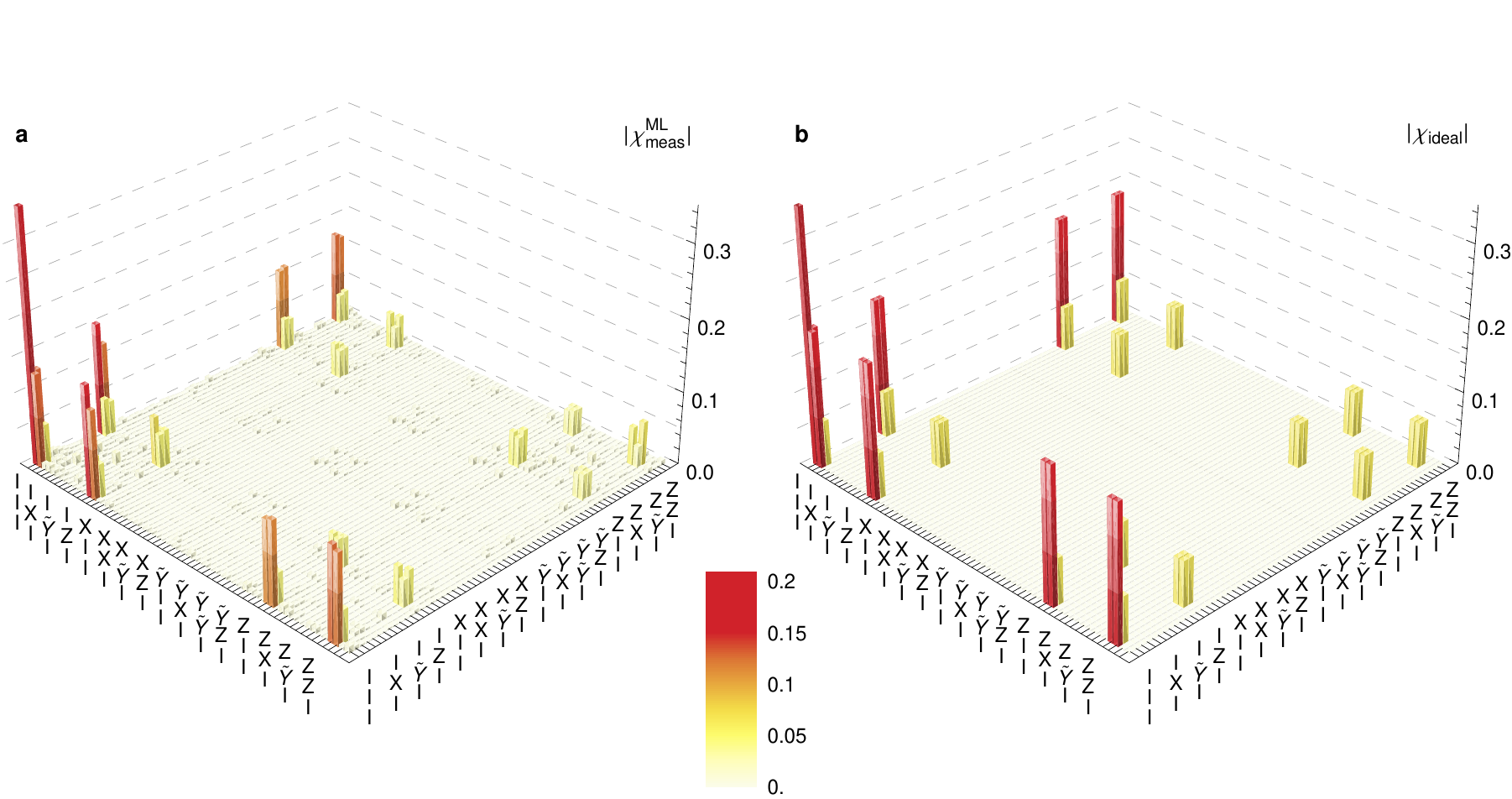}
  \caption{{\bf Process tomography of the Toffoli gate.} Bar chart of the absolute value of the measured  process matrix $\chi_\mathrm{exp}^\mathrm{ML}$ ({\bf a}) and the ideal process matrix  $\chi_\mathrm{ideal}$ ({\bf b}). The elements are displayed in the operator basis $\{III, IIX, II\tilde{Y}, \ldots ZZZ\}$, where $I$ is the identity and $\{X, \tilde{Y}, Z\}$ are the Pauli operators $\{\sigma_x, -i\sigma_y, \sigma_z\}$. The fidelity of the process matrix is $F = \mathrm{Tr}[\chi_\mathrm{exp}^\mathrm{ML}\chi_\mathrm{ideal} ] = 69\%$. The process fidelity estimated using the Monte Carlo certification method is $68.5\pm 0.5$\%.}
  \label{fig:chimatrix}
\end{figure*}

As an essential addition to the classical characterization of the gate by the truth table, we have performed full, three-qubit process tomography and reconstructed the process matrix, $\chi_\mathrm{exp}$, to characterize the quantum features of the Toffoli gate completely, overcoming the limited characterization provided by measurements of the phase fidelity only~\cite{Mariantoni2011a}. For this purpose, we prepared a complete set of 64 distinct input states by applying all combinations of single-qubit operations chosen from the set  \{id, $\pi/2_x$, $\pi/2_y$, $\pi_x$\} for each qubit, and performed state tomography on the respective output states.
The process matrix reconstructed directly from the data has a fidelity of $F = \mathrm{Tr}[\chi_\mathrm{exp}\chi_\mathrm{ideal} ] = 70\pm3\%$ (the error represents a $90\%$ confidence interval), where $\chi_\mathrm{ideal}$ is the ideal process matrix. Using a maximum-likelihood procedure~\cite{Jezek2003} to correct for unphysical properties of  $\chi_\mathrm{exp}$, we find that the obtained process matrix, $\chi_\mathrm{exp}^\mathrm{ML}$, has a fidelity of $F = \mathrm{Tr}[\chi_\mathrm{exp}^\mathrm{ML}\chi_\mathrm{ideal} ] = 69 \, \%$ with expected errors at the level of $3\,\%$. In Fig.~3a, $\chi_\mathrm{exp}$ shows the same key features as $\chi_\mathrm{ideal}$ (Fig.~3b).

To gain an accurate alternative estimate of the process fidelity without resorting to a maximum-likelihood procedure, we implemented Monte Carlo process certification following the steps described in ref.~\onlinecite{Silva2011}. First we define a Pauli observable as $\hat P_n = \prod_{j = 1,...,6}^{\otimes} \hat p_{n,j}$, a product of six single-qubit operators chosen from the set of the identity and the Pauli operators ($\hat p_{n,j}\in \{\mathbbm{1},\sigma_x, \sigma_y, \sigma_z\}$). Then we  determine  the 232 observables with non-vanishing expectation values $P_n = \mathrm{Tr} [\hat\rho_T \hat P_n]\neq 0$, where $\hat\rho_T$ is the Choi matrix of the Toffoli process. For each $\hat P_n$  we prepare all ($2^3 = 8$) eigenstates of the product of the first three operators comprising $\hat P_n$, apply the Toffoli operation to these states and measure the expectation value of the product of the last three operators in $\hat P_n$. Averaging over the results obtained with all eigenstates provides an estimate of $P_n$. Extracting all 232 expectation values in this way allows us to estimate the fidelity of the Toffoli gate as $68.5 \pm 0.5$\% using Monte Carlo process certification, which is in good agreement with the fidelity evaluated using tomography.

The scheme that we use to implement the Toffoli gate is generic and  can readily be applied to other systems because the majority of the quantum systems used as qubits have additional energy levels at their disposal. Reduction of the total gate time by use of qubit-qutrit gates together with the recent advances in the extension of the coherence times of the superconducting circuits~\cite{Bylander2011,Paik2011} indicates a path towards the  realization of practical quantum error correction.

\begin{acknowledgments}
We thank Stefan Filipp, Alexandre Blais for useful discussions and Kiryl Pakrouski for his contributions in early stages of the experimental work. This work was supported by the Swiss National Science Foundation (SNF), the EU IP SOLID and ETH Zurich.
\end{acknowledgments}

\end{document}